\documentclass[journal]{IEEEtran}

\ifCLASSINFOpdf
\else

\fi
\usepackage{graphicx}
\usepackage{epstopdf}
\usepackage{multicol}
\usepackage{url}
\usepackage{amsmath}
\usepackage{amssymb}
\usepackage{esint}
\usepackage{amsfonts}
\usepackage{multirow}
\usepackage{wrapfig}
\usepackage{cite}
\usepackage{subcaption}
\usepackage{lipsum}
\usepackage{float}	
\usepackage{algorithm}
\usepackage{algorithmic}
\usepackage{upgreek}
\usepackage[none]{hyphenat}
\usepackage[usenames,dvipsnames]{color}

\usepackage{geometry}
\begin{document}
\setlength{\abovedisplayskip}{3pt}
\setlength{\belowdisplayskip}{3pt}

\title{Impact of 3D Antenna Radiation Patterns on TDOA-Based Wireless Localization of UAVs}
\author{
\IEEEauthorblockN{Priyanka Sinha, Yavuz Yapici, and Ismail Guvenc}\\
\IEEEauthorblockA{Department of Electrical and Computer Engineering, North Carolina State University, Raleigh, NC}
\footnote[10]{This work has been supported in part by NASA under the Federal Award NX17AJ94A.}\\
Email: {\tt\{psinha2,yyapici,iguvenc\}@ncsu.edu}
\thanks{This work has been supported in part by NASA under the Federal Award number NX17AJ94A.}
}
\maketitle
\begin{abstract}
Next big commercial applications of drones
require to fly the drone beyond the visual line of sight (BVLOS). This inevitable ability to fly BVLOS will also necessitate the ability to keep track of the drone's location, in order to ensure successful completion of the intended service. In this context, we explore the fundamental limits of 3D localization of drones in conjunction with the effects of the 3D antenna radiation patterns. Although localization of drone/unmanned aerial vehicle (UAV) is a well-studied topic in the literature, its relationship to the antenna effects remains mostly unexplored. In this paper, we investigate the impact of antenna radiation pattern on the accuracy of time-difference-of-arrival (TDOA)-based localization of the UAV. Specifically, we consider a scenario where a fixed number of radio-frequency (RF) sensors, placed at some known locations on the ground, collect the TDOA measurements from the signals transmitted from the UAV, and estimate the location of the UAV from these observations. In order to study the impact of the antenna effects on the fundamental limits of the TDOA-based positioning scheme, we develop a simple analytical model to approximate the total antenna gains experienced by an air-to-ground (A2G) link, for various orientations of the antennas. We then derive the Cramer-Rao lower bound for the TDOA based localization scheme, for all three combinations of the transmit and the receive antenna orientations: vertical-vertical (VV), horizontal-horizontal (HH), and vertical-horizontal (VH).  
\end{abstract}
\begin{IEEEkeywords}
Antenna effects, cellular positioning, Cramer-Rao lower bound (CRLB), drone-localization, TDOA based localization, unauthorized drone detection.
\end{IEEEkeywords}
\vspace{-4mm}

\section{Introduction}
\label{sect:Introduction}
Drone localization has a wide range of applications 
in military, commercial, government, and recreational scenarios. Depending on the context, the drone may or may not participate in the localization process, either because the drone communication system might not be designed for positioning purposes, or it may be a non-cooperating and potentially malicious drone~\cite{7470933,guvenc2018detection,azari2018key}. There has been some recent studies in the literature to use the radio frequency (RF) signals from drones for the purpose of detection, classification, localization, and tracking of drones, regardless of whether the drone is cooperating or not~\cite{whitepaper1,whitepaper2,guvenc2018detection,azari2018key}.

\par There are several parameters associated with an RF signal that can be measured and used to localize of the source of the signal, such as the received signal strength (RSS), angle-of-arrival (AOA), time-of-arrival (TOA), and time-difference-of-arrival (TDOA)~\cite{Gezici:2008:SWP:1341571.1341575}. Due to the simplicity of implementation, localization schemes involving RSS, TOA, and TDOA are more common as compared to the AOA methods, and the time based methods are usually preferred over the RSS based methods due to the higher degree of accuracy they provide. Between the TOA and the TDOA based approaches, TDOA  is more suitable to a wider range of general applications, as it does not require the source clock to be synchronized with that of the RF sensor.

Due to their advantages, the TDOA based localization schemes have been studied extensively in the literature. In~\cite{5208736}, the authors provide a survey of the existing time of arrival (TOA) based algorithms, provide associated Cramer-Rao lower bound (CRLB) for different approaches, and review non-line-of-sight (NLOS) mitigation techniques in the literature. In~\cite{1212671}, the authors study the fundamental limits of TOA-based localization scheme for wireless sensor networks. In~\cite{6965725}, TDOA-based method is introduced for \emph{direct position estimation} in a 2D plane and the associated CRLB are derived. 
In spite of the large volume of the existing literature, a study of the TDOA based localization that takes the effect of antenna radiation patterns into account is lacking in the literature to our best knowledge. The effect of the antenna patterns on the air-to-ground (A2G) channels is investigated in~\cite{2018arXiv181001442C}, and the impact of the antenna patterns on the received signal strength at the ground units is demonstrated. 
\begin{figure}[t]\vspace{-1mm}
	\centering	\includegraphics[width=0.8\columnwidth]{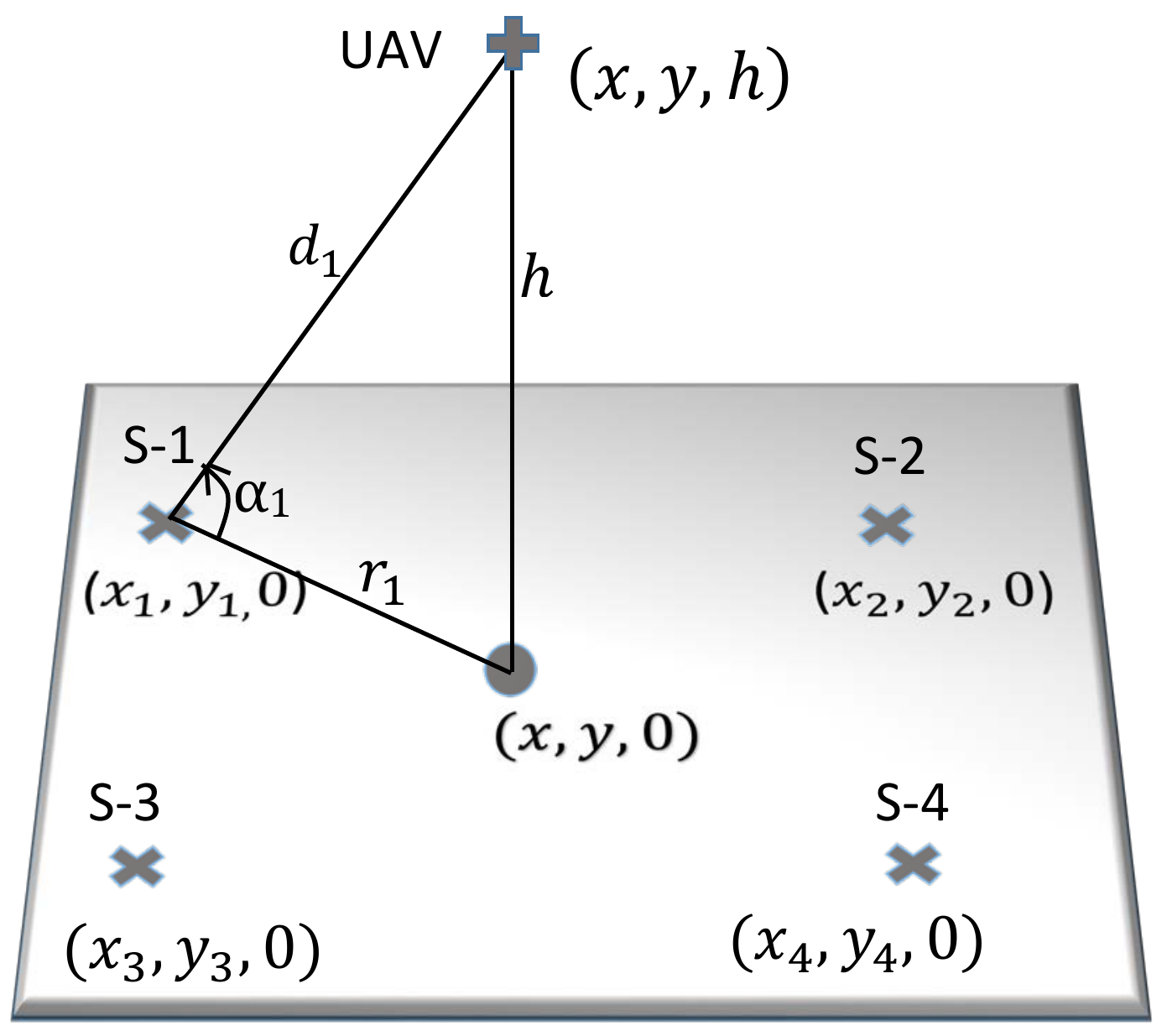}\vspace{1mm}
	\caption{Reference sensors with fixed locations, that collect TOA measurements for drone localization.}
	\label{fig:SystemModel11}
    \vspace{-6mm}
\end{figure}

In this paper, we consider a simple antenna model as used in~\cite{2018arXiv181001442C}, 
and we study the impact of the antenna alignment on the fundamental limits of the 3D localization of a UAV. To the best of our knowledge, such work is not a part of the literature yet. The main contributions of our work are as follows: 1) a simple yet effective analytical model is introduced to approximate the antenna gains in the A2G link for the different orientations of the transmit and the receive antennas; 2) closed form expressions are derived demonstrating the impact of antenna gains on the variance of the TDOA measurement noise, assuming the availability of Line-of-sight (LOS) A2G links; 3)  the CRLB on the TDOA based localization is derived; and finally 4) the performance of the localization scheme for the different antenna patterns are characterized, and computer simulations are used to evaluate localization coverage and accuracy metrics for various antenna configurations. 

\section{System model}
\label{sect:system_model}

\subsection{Extraction of the TOA from RF Signals}

The TDOA measurement between a pair of time synchronized RF sensors is obtained by calculating the difference between the TOA estimates measured at each of these sensors. Subtracting one TOA estimate from another eliminates the dependence of the estimated distance between the UAV and the sensor, on the unknown time of transmission at the UAV. Assuming that an efficient maximum-likelihood (ML) estimator is used at the sensors to obtain the TOA estimates, the TOA measurement noise at the sensors can be modelled as zero-mean Gaussian noise, due to the asymptotic normality property of ML estimators. We consider a system model where we have a fixed number of ground RF-sensors or fixed terminals (FTs) with known locations $(x_i, y_i)$, to collect the TOA measurements from the signals arriving from a UAV located at $(x, y, h)$.  Fig.~\ref{fig:SystemModel11} illustrates such a model with four ground FTs, where it is desired to estimate the 3-D location coordinates $(x, y, h)$ of the UAV. 

Let the 3D distance between the UAV and the $i^{\rm th}$ sensor, S-$i$,  be denoted as $d_i$, and let the actual (measured) value of the TOA be denoted as $\tau_i$ ($\hat{\tau}_i$). Then the relationship among these quantities are given as:
\begin{align}
{d}_i&=\sqrt[]{{\left(x_i-x\right)}^2+{\left(y_i-y\right)}^2+h^2}~,\label{eq:dist}\\
{\tau}_i&=\frac{d_i}{c} + t_0~,\label{eq:TOF}\\
{\hat{\tau}}_i&=\tau_i + n_i~,\label{eq:TOA}
\end{align}


where ${\rm n_1}\sim\mathcal{N}(0,\sigma^2_{i})$, is the additive white Gaussian measurement noise, and $c$ is the speed of propagation.
The TOA estimates in \eqref{eq:TOA} are conventionally obtained from correlating receivers.
We assume that each of the sensors intercepts the UAV signal within the time interval $(0, T)$, where the UAV signal is assumed to have a known waveform $s(t)$ that is transmitted for a known duration of $T_s \leq T$, starting from an unknown time $t_0 \le T$. As it is not unusual to have a dominant line-of-sight (LOS) path for A2G channels, the signal received at the $i^{\rm th}$ sensor, in presence of additive white Gaussian noise (AWGN) can be given as:
\begin{align}
r_{i}(t) = {A_i}s(t-\tau_{i})+n(t)~,\label{eq:TOA CORR}
\end{align}
where $\tau_i$ is the true value of the TOA at the $i^{\rm th}$ BS, $A_i$ is the amplitude factor representing large scale fading, and $n(t)$ is the white Gaussian noise with zero mean and spectral density $\frac{N_{0}}{2}$. Please note that, due to the assumption of having a strong LOS component, we ignore the impact of Doppler spread, as the number of multipath components and the angular spread of the incoming waves are expected to decrease, which in turn reduces the Doppler spread~\cite{a2gsurvey}.  
It is desired to estimate the time of arrival, $\tau_i$ in \eqref{eq:TOA CORR}, and for this continuous time model, the log-likelihood function~\cite{6965725} for the $i^{\rm th}$ sensor is given as:
\begin{align}
l(\tau_i) = {-\frac{1}{N_0}}{\int_{0}^{T} ({r_i}(t)-{A_i}s(t-\tau_i))^2} dt.
\end{align}

Since we argue that the above ML estimator achieves asymptotic optimality in a LOS-dominated environment, we conclude that the covariance of the TOA estimates at all the sensors attains the CRLB, and is given as the inverse of the Fisher Information Matrix (FIM), $\boldsymbol{I}(\boldsymbol{\tau})$. According to~\cite{Estimation1993Kay}, the corresponding FIM is obtained as:
$\boldsymbol{I}(\boldsymbol{\tau}) =\frac{\partial{{\boldsymbol{r}^T}(\tau)}}{\partial\tau}{{\Sigma}^{-1}}\frac{\partial{{\boldsymbol{r}}(\tau)}}{\partial\tau}~,$
where $\boldsymbol{r} = [r_1, r_2, r_3, .. r_{\small{N}}]^T$ is the received signal vector and $\boldsymbol{\Sigma}$ is the covariance of the AWGN at the sensors. Since we assume independent and identically distributed (iid) noise samples, $\boldsymbol{\Sigma}$ becomes a diagonal matrix ${\sigma}^2\boldsymbol{I}_{N}$, and inverse FIM is
\begin{align}
\boldsymbol{I}^{-1}(\boldsymbol{\tau})&=\left(
\begin{array}{cccc}
\sigma^2_{1} & 0 & \cdots & 0 \\
0 & \sigma^2_{2} & \cdots & 0 \\
\vdots & \vdots & \ddots &  \vdots  \\
0 & 0 & \cdots & \sigma^2_{N}
\end{array}
\right)~,\label{eq:TOA_noise}
\end{align}

Due to the efficient estimator assumption, the TOA measurement noise at the $i^{\rm th}$ sensor, in \eqref{eq:TOA}, becomes a zero-mean Gaussian random variable with variance $\sigma^2_{i} = [{\boldsymbol{I}^{-1}}(\boldsymbol{\tau})]_{ii}$, which can also be written as~\cite{Gezici:2008:SWP:1341571.1341575}:
\begin{align}
\sigma_{i} &= \frac{1}{2\sqrt{2}\pi{\sqrt{SNR_i}\beta}}= \frac{k}{SNR_i}~,\label{eq:TOA CRLB}
\end{align}
where $\beta$ is the effective drone signal bandwidth, and $SNR_i=\frac{{{A_i}^2}{E_S}}{N_0}$, with $E_s$ representing the energy in the transmit signal $s(t)$.
For the broader objective of this paper
we summarize the impact of factors other than signal-to-noise ratio (SNR) on the TOA measurement noise, using a proportionality constant $k$.
\begin{figure}[t]
	\centering
	\includegraphics[width=0.8\columnwidth]{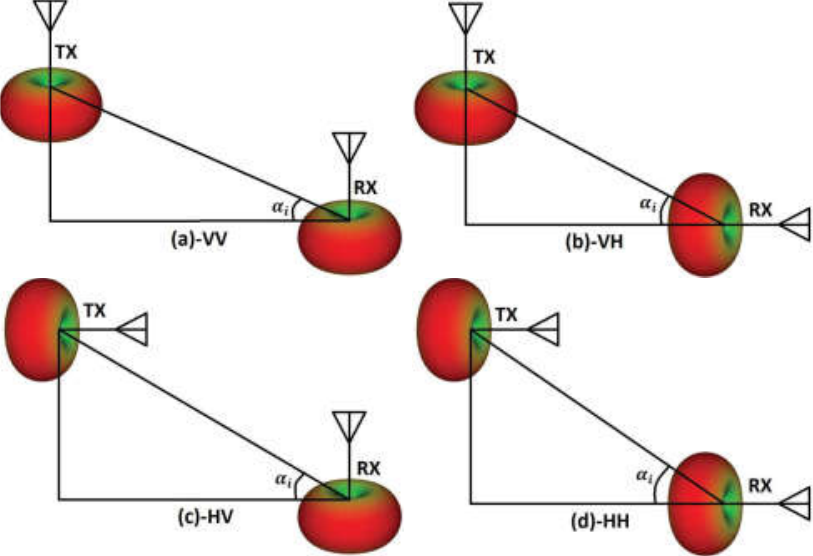}
	\caption{Analytical model for the antenna gains considering vertical-vertical antenna configuration.}
	\label{fig:antenna_model}
    \vspace{-7mm}
\end{figure}

\subsection{Analytical Model for Antenna Pattern}

In order to study the impact of the antenna radiation patterns on the TOA CRLB, we characterize the relationship between the total antenna gain experienced at the $i^{\rm th}$ sensor, and the variance of the TOA measurement noise. In this paper, we use a simple 
antenna gain function 
to characterize the gain from dipole antenna radiation patterns~\cite{2018arXiv181001442C}. In particular, we approximate the antenna gain as a sine/cosine function of the elevation angle ($\alpha_i$) between the aerial transmitter (the UAV), and the receiver at the $i^{\rm th}$ ground sensor depending on the orientation of the transmit and receive antennas. We consider $4$ such antenna orientations, namely, the VV pattern, where both Tx and Rx antennas are oriented vertically, the HH pattern, where both the transmit and the receive antennas are oriented horizontally, and the VH/HV pattern, where the transmit and the receive antennas are respectively, vertically and horizontally and vertically and horizontally oriented. 

\begin{table}[h!]
\centering
 \caption{Total antenna gain for different antenna patterns}\label{Tab1A}
 \begin{tabular}{|c||c|c|c|} 
 \hline  & {\bf VV} & {\bf VH} & {\bf HH} \\ 
 \hline  $G_{\rm RX}(\alpha_i)$ & $\cos(\alpha_i)$ & $\cos(\alpha_i)$ & $\sin(\alpha_i)$ \\ 
 \hline  $G_{\rm TX}(\alpha_i)$ & $\cos(\alpha_i)$ & $\sin(\alpha_i)$ & $\sin(\alpha_i)$ \\ 
 \hline  $G_{\rm RX}(\alpha_i)G_{\rm TX}(\alpha_i)$ & $\cos^2(\alpha_i)$ & $0.5\sin(2\alpha_i)$ & $\sin^2(\alpha_i)$ \\
\hline
 \end{tabular}
\end{table}

Fig.~\ref{fig:antenna_model}(a) and Fig.~\ref{fig:antenna_model}(b) show the VV and the HH scenario for antenna radiation patterns, respectively, where the doughnut-like radiation pattern in the 3D plane are also illustrated. The VH and the HV patterns are identical in terms of the total antenna gain, and are illustrated in Fig.~\ref{fig:antenna_model}(c) and Fig.~\ref{fig:antenna_model}(d), respectively. The total antenna gain experienced at the receiver, $G_{\rm TX}G_{\rm RX}$, for the different scenarios are summarized in Table~\ref{Tab1A}.

As a result, we can represent the RSS at the $i^{\rm th}$ ground receiver from an aerial transmitter as: 
\begin{equation}
P_{\rm RX}(\alpha_i) = P_{\rm TX}G_{\rm TX}(\alpha_i)G_{\rm RX}(\alpha_i)\left(\frac{\lambda}{4\pi d_i}\right)^{\gamma}~,\label{Eq_1}
\end{equation}
where $P_{\rm TX}$ is the transmitted signal power at the UAV, $G_{\rm TX}(\alpha_i)\leq 1$ is the transmitter antenna gain, $G_{\rm RX}(\alpha_i)\leq 1$ is the receiver antenna gain, $\alpha_i$ and $d_i$ are the elevation angle and the 3-D distance between the transmitter and the receiver, respectively, $\lambda$ is the  wavelength, and $\gamma$ is the path loss exponent, which is assumed as two in the rest of this paper due to LOS assumption. With help of \eqref{Eq_1}, we can now represent the receive SNR as a function of the antenna gains:
\begin{align}
SNR_i = \frac{P_{\rm RX}(\alpha_i)}{{{\sigma}_{\rm n}}^2}={\left(\frac{\lambda}{4\pi}\right)}^2\frac{{P_{\rm TX}}{G_{\rm TX}(\alpha_i)}{G_{\rm RX}(\alpha_i)}}{{{\sigma_{\rm n}}^2}{d_i}^2}~,\label{eq:time domain SNR}
\end{align}

where ${\sigma_{\rm n}}^2$ is the variance of the thermal AWGN. Using \eqref{eq:TOA CRLB}, the relationship between the total antenna gain, and the variance of the TOA measurement noise at the $i^{\rm th}$ sensor is: 
\begin{align}
{\sigma_i}^2 =\frac{{k_0}{\sigma_{\rm n}}^2{d_i}^2}{{P_{\rm TX}}{G_{\rm TX}(\alpha_i)}{G_{\rm RX}(\alpha_i)}}~,\label{eq:toa_var_gain}
\end{align}
where the proportionality constant $k_0$ summarizes the impact of all factors other than the total antenna gain, AWGN noise variance and the 3-D distance between the UAV and the $i^{\rm th}$ sensor, and is given as~\cite{Zekavat:2011:HPL:2161958}:
$k_0 = \frac{2Ef^2}{{\beta}^2c^2}~,$
where $\beta$ is the bandwidth of the drone signal, $E$ is the energy in the drone signal at the UAV transmitter, $f$ is the center operating frequency of the ground sensors, and $c$ is the speed of propagation. Using the expression for the elevation angle $\alpha_i=\tan^{-1}{\left(\frac{h}{\sqrt{(x-x_i)^2 + (y-y_i)^2}}\right)}$, in~\eqref{eq:toa_var_gain}, the relationship between the variance of the TOA measurement noise ${\sigma_i}^2$, and the unknown location of the UAV ($x,y,h$), can be represented as follows:\\
\begin{align}
{\sigma_{\text{i-\tiny{VV}}}}^2&=\frac{k_0{{\sigma_{\rm n}}^2}\left(\left(x-x_i\right)^2 + \left(y-y_i\right)^2+ h^2\right)^2}{\left(x-x_i\right)^2 + \left(y-y_i\right)^2}~,\label{eq:var_toa_vv}\\
{\sigma_{\text{i-\tiny{HH}}}}^2&=\frac{k_0{{\sigma_{\rm n}}^2}\left(\left(x-x_i\right)^2 + \left(y-y_i\right)^2+ h^2\right)^2}{h^2}~,\label{eq:var_toa_hh}\\
{\sigma_{\text{i-\tiny{VH}}}}^2 & = 
\frac{k_0{{\sigma_{\rm n}}^2}\left(\left(x-x_i\right)^2 + \left(y-y_i\right)^2+ h^2\right)^2}
{ h\sqrt{\left(x-x_i\right)^2 + \left(y-y_i\right)^2} }
~,\label{eq:var_toa_vh}
\end{align}
\begin{align}
{\sigma_{\text{i-\tiny{U}}}}^2&=k_0{{\sigma_{\rm n}}^2}\left(\left(x-x_i\right)^2 + \left(y-y_i\right)^2+ h^2\right)~,\label{eq:var_toa_uniform}
\end{align}
\\
where ${\sigma^2_{\text{i-VV}}}$, ${\sigma^2_{\text{i-HH}}}$, ${\sigma^2_{\text{i-VH}}}$, and ${\sigma^2_{\text{i-U}}}$, are respectively the variance of the TOA measurement noise for the VV, HH, VH, and `Uniform/Isotropic' antenna patterns at the $i^{\rm th}$ sensor.

\subsection{Extraction of the TDOA from Measurements}

As discussed in Section~\ref{sect:system_model}, the TDOA observations are obtained by computing the difference between two TOA estimates. In this paper, we use a common reference sensor, with respect to which all the TDOA measurements are made. 
We choose $S$-$1$ in Fig.~\ref{fig:SystemModel11}, as the reference sensor, and hence, with $N$ sensors, we obtain a TDOA measurement vector with $N-1$ elements. The true TDOA and its measured value are then given by $\tau_{1i}=\frac{d_1}{c}-\frac{d_i}{c}$ and $\quad \hat\tau_{1i}=\hat{\tau_1}-\hat{\tau_i}$, respectively.
Thus $\hat\tau_{1i}$ is a Gaussian random variable ${\rm \hat\tau_{1i}}\sim\mathcal{N}\left(\frac{d_1-d_i}{c},\sigma^2_{1}+\sigma^2_{i}\right)$. Due to having a common reference sensor, the TDOA measurements are not independent, and the dependence is modeled through the covariance matrix $\boldsymbol{R}\left(\boldsymbol{\rm x}\right)$. Thus the TDOA measurement vector $\boldsymbol{z}=\left[\hat\tau_{12},\hat\tau_{13} ... \hat\tau_{1N}\right]$ is jointly-Gaussian distributed
${\rm \boldsymbol{z}}\sim\mathcal{N}\left(\boldsymbol{\mu}\left(\boldsymbol{\rm x}\right),\boldsymbol{R\left(\boldsymbol{\rm x}\right)}\right)$.
The mean vector and the covariance matrix have been parameterized to show their dependence on the unknown location of the UAV, $\boldsymbol{\rm x}=(x,y,h)$, and they can be respectively written as follows:
\begin{align}
\boldsymbol{\mu}(\boldsymbol{\rm x}) &= \frac{1}{c}{\bigg[\left(d_1-d_2\right), \left(d_1-d_3\right) ... \left(d_1-d_N\right)\bigg]}~\label{eq:mean_vec},\\
\boldsymbol{R}(\boldsymbol{\rm x}) &= \left(
\begin{array}{cccc}
\sigma^2_{1}+\sigma^2_{2} & \sigma^2_{1} & \vdots & \sigma^2_{1} \\
\sigma^2_{1} & \sigma^2_{1}+\sigma^2_{3} & \vdots & \sigma^2_{1} \\
\vdots & \vdots & \ddots & \vdots \\
\sigma^2_{1} & \sigma^2_{1} & \cdots & \sigma^2_{1}+\sigma^2_{(N-1)}
\end{array}
\right)~.\label{eq:cov_matrix}
\end{align}
\section{Localization CRLB with Antenna Effects}
\label{sect:CRLB}
Given~\eqref{eq:mean_vec} and \eqref{eq:cov_matrix}, the likelihood function for the TDOA measurement vector, $\boldsymbol{z}$ can be written as a function of the UAV location $(x,y,h)$ as follows:
\begin{align}
    p\left(\boldsymbol{z}|\boldsymbol{x}\right) = \frac{\exp\left(\frac{-\left(\boldsymbol{z}-\boldsymbol{\mu}\left(\boldsymbol{x}\right)\right){\boldsymbol{R}^{-1}\left(\boldsymbol{x}\right)}\left(\boldsymbol{z}-\boldsymbol{\mu}\left(\boldsymbol{x}\right)\right)^{T}}{2}\right)}{{{{\sqrt{2\pi}}|\boldsymbol{R}\left(\boldsymbol{x}\right)|}}}~,
\end{align}
and therefore, the ML estimate of $\boldsymbol{x}$ is given by:
\begin{align}
\boldsymbol{\hat{\boldsymbol{x}}}=\arg \max_{\mbox{$\boldsymbol{\boldsymbol{x}}$}}~~ p\left(\boldsymbol{z}|\boldsymbol{x}\right)~.
\end{align}
In our model both the mean $\boldsymbol\mu(\boldsymbol{x})$, and the covariance $\boldsymbol R(\boldsymbol{x})$ of the PDF is dependent on the unknown UE location $\boldsymbol{x}=(x,y,h)$. Therefore, the relationship between the covariance matrix $\boldsymbol{c}_{\boldsymbol{\hat{x}}}$ of an unbiased vector estimator $\boldsymbol{\hat{x}}$ and the Fisher information matrix (FIM), $I_{\boldsymbol{\hat{x}}}$ can be given as~\cite{Estimation1993Kay}:
\begin{align}
\boldsymbol{c}_{\boldsymbol{\hat{x}}} - \begin{small}\boldsymbol{{I}_{\boldsymbol{\hat{x}}}^{-1}} \ge \boldsymbol{0}\end{small},\label{eq:covariance matrix}
\end{align}
where, for $i=0,1,2$, and $j=0,1,2$, the $(i,j)$ element of the FIM is given as: 
\begin{align}
    I_{ij} &= {\frac{\partial \boldsymbol{\mu}(\boldsymbol{x})}{\partial x_i}}{\boldsymbol{R}^{-1}(\boldsymbol{x})}{\frac{\partial \boldsymbol{\mu}(\boldsymbol{x})}{\partial x_i}}^{T}\nonumber\\
    &+ \frac{1}{2}\text{trace}\big({\boldsymbol{R}^{-1}(\boldsymbol{x})}\frac{\partial \boldsymbol{R}(\boldsymbol{x})}{\partial x_i}{\boldsymbol{R}^{-1}(\boldsymbol{x})}\frac{\partial \boldsymbol{R}(\boldsymbol{x})}{\partial x_j}\big)~,\label{eq:FIM_ij}
\end{align}
where,
\begin{align}
\frac{\partial{\boldsymbol{\mu}}}{\partial x_i}\left(\boldsymbol{\rm x}\right) &= {\frac{1}{c}}{\left[\frac{\partial{\left(d_1-d_2\right)}}{\partial x_i}, \frac{\partial{\left(d_1-d_3\right)}}{\partial x_i} \cdots \frac{\partial{\left(d_1-d_N\right)}}{\partial x_i}\right]}~,\nonumber\\
{\frac{\partial \boldsymbol R\left(\boldsymbol{\rm x}\right)}{\partial x_i}} &= \left(
\begin{array}{cccc}
\frac{\partial\left({\sigma^2_{1}+\sigma^2_{2}}\right)}{\partial x_i} & \frac{\partial{\sigma^2_{1}}}{\partial x_i} & \cdots & \frac{\partial{\sigma^2_{1}}}{\partial x_i} \\
\frac{\partial{\sigma^2_{1}}}{\partial x_i} & \frac{\partial{\left(\sigma^2_{1}+\sigma^2_{3}\right)}}{\partial x_i} & \cdots & \frac{\partial{\sigma^2_{1}}}{\partial x_i} \\
\vdots & \vdots & \ddots & \vdots  \\
\frac{\partial{\sigma^2_{1}}}{\partial x_i} & \frac{\partial{\sigma^2_{1}}}{\partial x_i} & \cdots & \frac{\partial{\left(\sigma^2_{1}+\sigma^2_{M}\right)}}{\partial x_i}
\end{array}
\right)~,\nonumber
\end{align}
with $M=(N-1)$. The partial derivatives in the above equations can be obtained using \eqref{eq:var_toa_vv} through \eqref{eq:var_toa_vh}, and \eqref{eq:mean_vec}. Thus, the mean square error (MSE) can be given written as:
\begin{align}
E(x,y,h) \geq {\text{trace}[{\boldsymbol{I}^{-1}}(x,y,h)]}~.\label{eq:RMSE CRLB}
\end{align}
\section{Localization Metrics}
\label{sect:metrics}
In this section we introduce a number of localization performance metrics that we use for the evaluation of the localization performance in the rest of this paper. First of all, we define the system area under the consideration, as a subset of the 2D Euclidean plane, $\mathcal{A} \subset \mathbb{R}^2$.
From the \eqref{eq:RMSE CRLB}, we see that the CRLB-RMSE for a drone location $(x,y,h)$, is a function of the true value of the parameters. Then for a constant drone height of $h$, $E$ can be defined as a mapping from $\mathcal{A}$ $\rightarrow{\mathbb{R}}$. The distribution of the random variable $E$, can then be written as
\begin{align}
    F_{E}(\delta)={\rm Pr}(E\le\delta)~.\label{eq:RMSE_dist}
\end{align}

\subsection{Coverage Metrics} For the purpose of this paper we use the Boolean notion of coverage, where the drone location $(x,y,h)$ is considered to be covered under the localization scheme, if the CRLB-RMSE at $(x,y,h)$ is less than a certain threshold $\delta$. Similarly, a sub-region of the system area, $\mathcal{A}_i \subset \mathcal{A}$, is considered to be covered under the localization scheme, if $\forall (x,y)\in \mathcal{A}_i$, the MSE, $E(x,y) \le \delta$. Thus we define a coverage function as:
\begin{align}
    {\rm Cov}(\mathcal{A}_i)= \begin{cases} 
      1, & \text{if~} \forall (x,y) \in \mathcal{A}_i, E(x,y)\le \delta \\
      0, & \text{Otherwise} 
   \end{cases}~.
\end{align}
We then consider the following two coverage metrics.

\paragraph{Area Coverage Efficiency (ACE)} The ACE is a measure of the aggregate localization coverage over the entire system area. We define ACE as the ratio of the area that is covered according to the Boolean notion of coverage to the entire system area. Thus the mathematical definition of ACE can be given as below:
\begin{align}
    \text{ACE}(\mathcal{A},\delta,\zeta)=\frac{\text{area}(\bigcup_{i:{Cov}(\mathcal{A}_i)=1}^{} \mathcal{A}_i)}{\text{area}(\mathcal{A})}~.\label{eq:ACE}
\end{align}
From the fundamental definition of probability, \eqref{eq:ACE} can be expressed in terms of the distribution function of $E$, as below:
\begin{align}
    \text{ACE}(\mathcal{A},\delta)=Pr(E\le\delta;\zeta)=F_{E}(\delta,\zeta)~.\label{eq:ACE_dist}
\end{align}
In the rest of this paper, we assume that the ACE is measured in conjunction with ${\delta}=100$~meters, and is named as the $100$-meters-error-bound-coverage\%. 

\paragraph{Quality of Localization (QOL)} As we can see, if the error tolerance threshold, $\delta$ is increased, due to the non-decreasing nature of a cumulative distribution function (CDF), the value of the ACE will also be higher, while on the other hand the QOL will be diluted. Thus, we use the required error tolerance threshold, $\delta$ to achieve the targeted $p$\% ACE, as a combined measure of the QOL and the ACE. The QOL can be mathematically defined as:
\begin{align}
    \text{QOL}(\mathcal{A},p,\zeta)={F_{E}}^{-1}(p,\zeta)~.\label{eq:QOL_dist}
\end{align}
In this paper, we use $p=80$\%, as the required ACE against which we measure the QOL, and the metric is termed as the $80$\%-coverage-error-bound.
\subsection{Accuracy Metrics} In this paper, we use a most commonly used measure of accuracy, the median RMSE. The median RMSE is defined as below:
\begin{align}
    \text{Median}(E)={F_{E}}^{-1}(0.5)~.\label{eq:median}
\end{align}
As we can see the median is the same as the QOL, subject to $50$\% coverage. However, we don't use the median as a measure of the QOL, as a requirement of $\text{ACE}=50$\%, is too low for any practical application.

\begin{figure*}[t]
\centering 
	\begin{multicols}{2}
    \begin{subfigure}{0.5\textwidth}
	\includegraphics[width=3in]{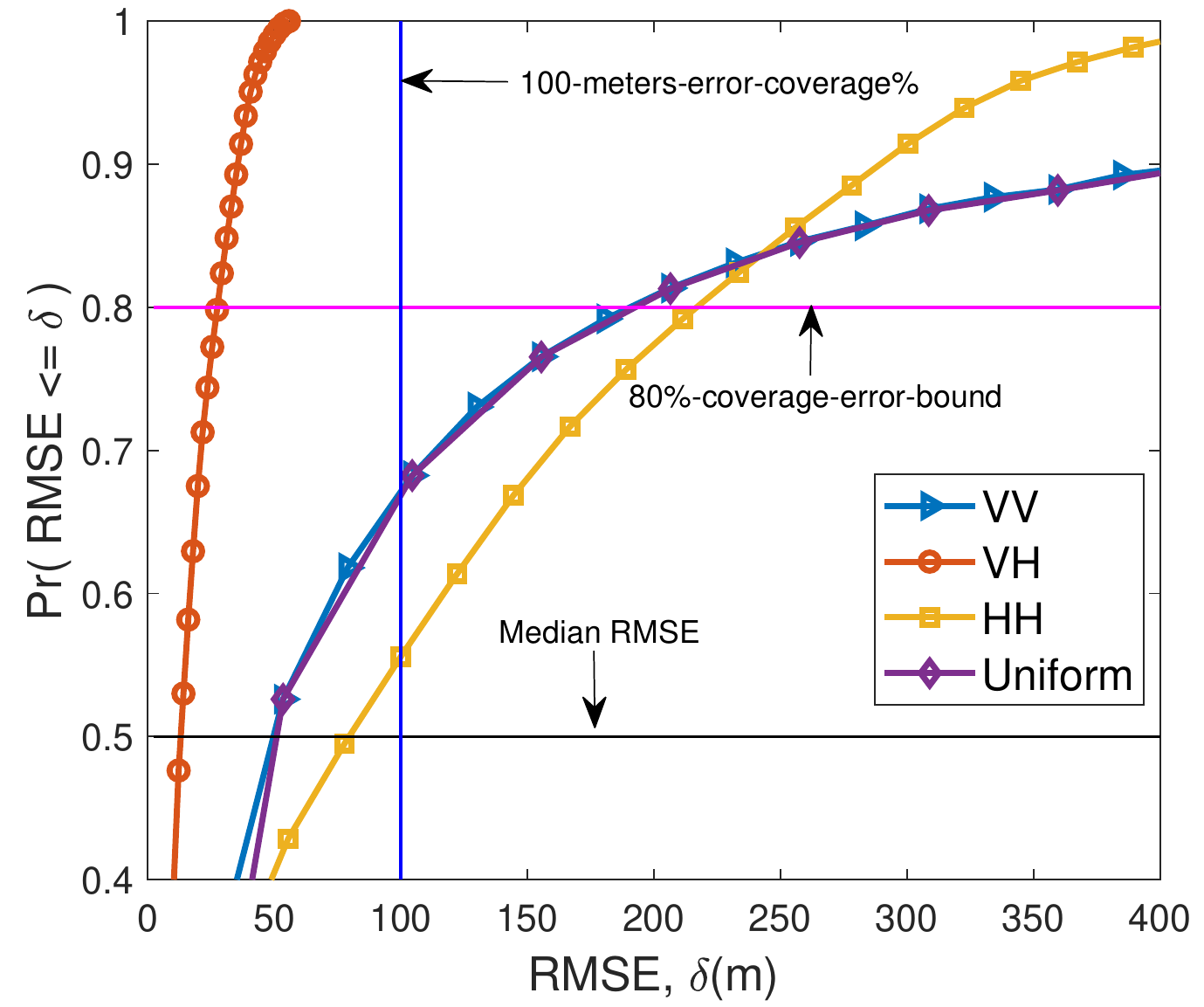} 
    \caption{Drone height: 10 meters.}
	\end{subfigure}
 
	\begin{subfigure}{0.5\textwidth}
    \includegraphics[width=3in]{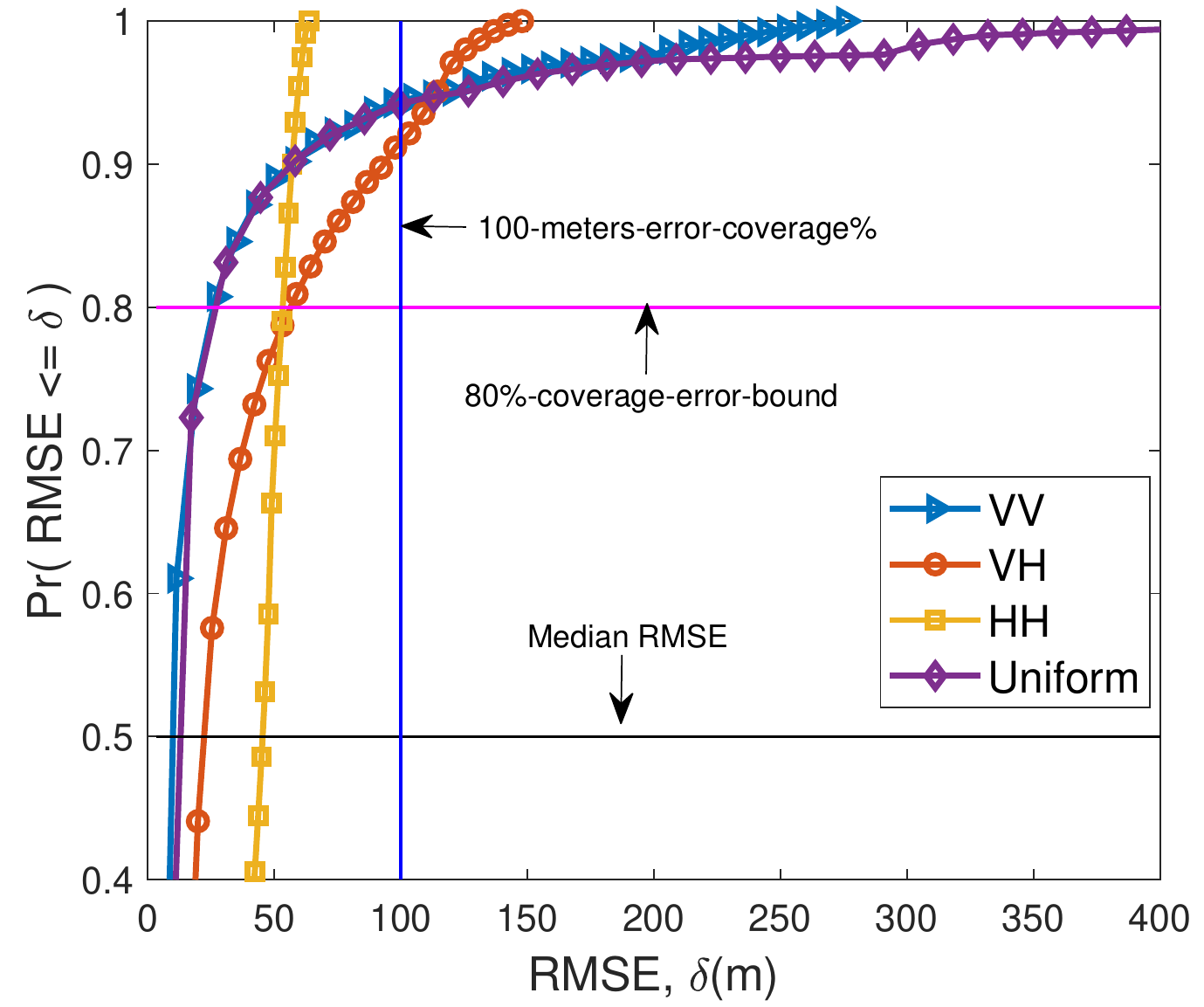} 
       \caption{Drone height: 100 meters.}
    \end{subfigure}
    \end{multicols} 
    
    \begin{multicols}{2}
    \begin{subfigure}{0.5\textwidth}
	\includegraphics[width=3in]{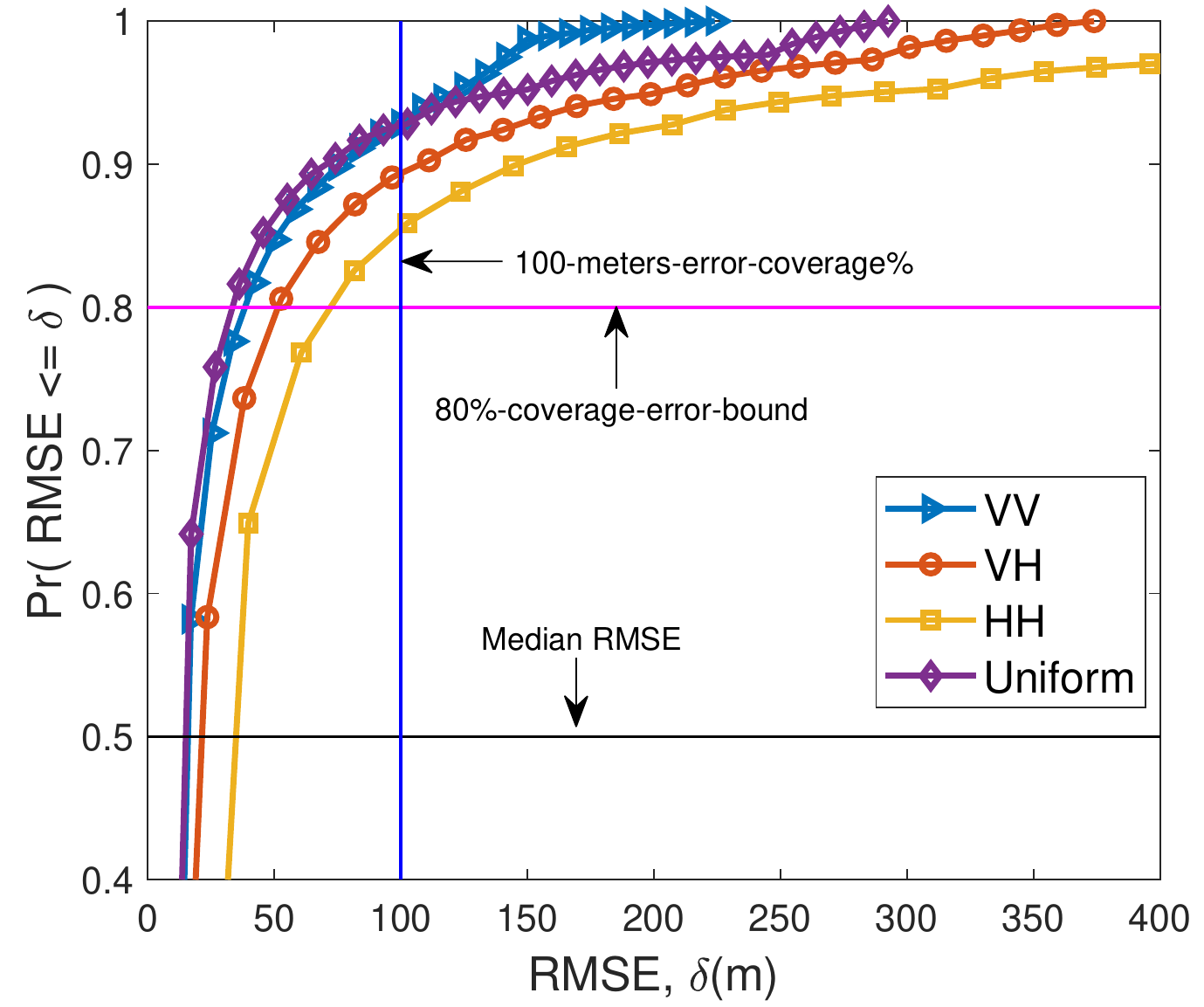} 
       \caption{Drone height: 300 meters.}
	\end{subfigure}

	\begin{subfigure}{0.5\textwidth}
    \includegraphics[width=3in]{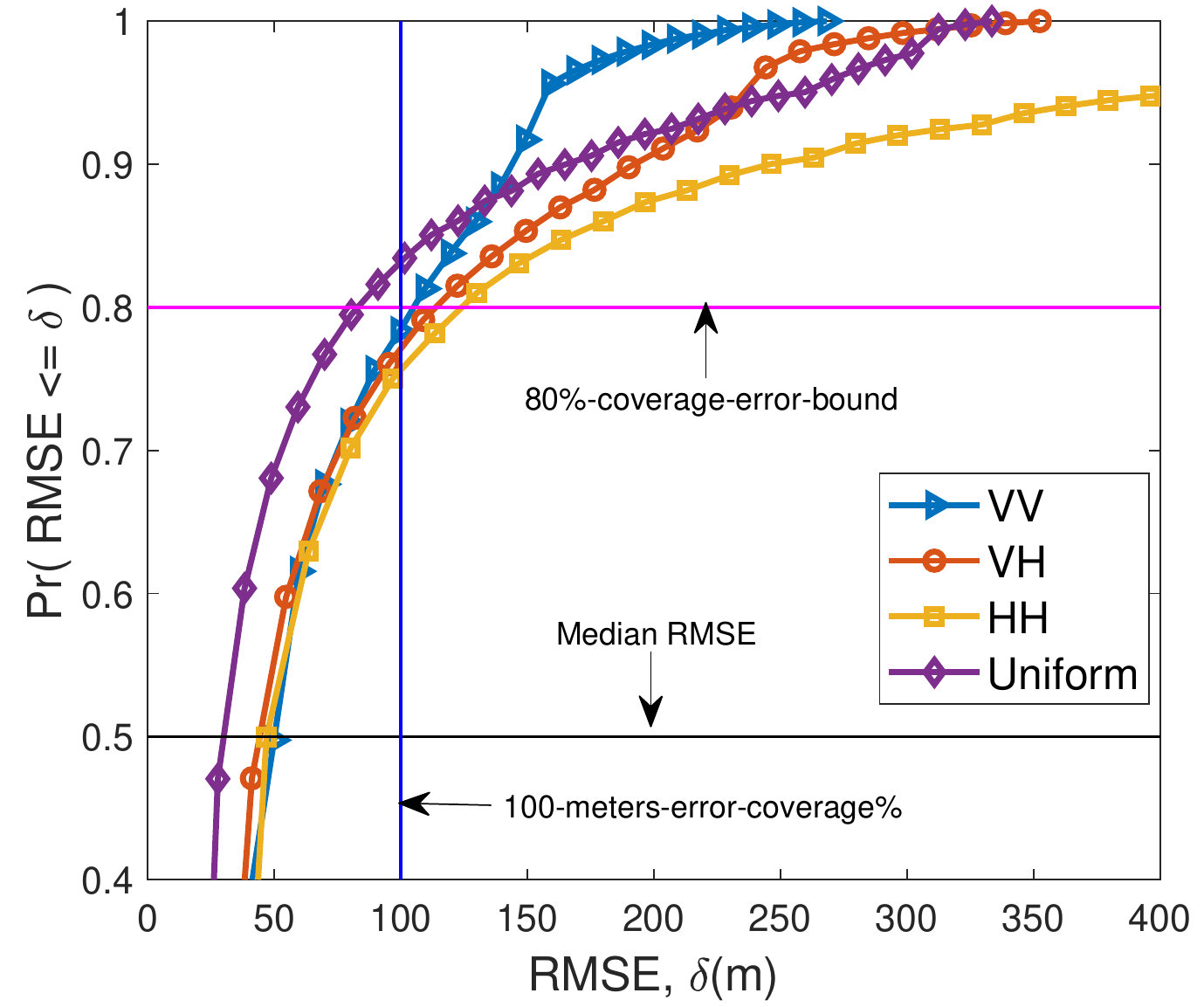} 
       \caption{Drone height: 500 meters.}
    \end{subfigure}
    \end{multicols} 
\vspace{-2mm}
\caption{Cumulative distributions of CRLB RMSE for various antenna orientations (VV, VH, HH, Uniform) at UAV altitudes of (a) 10 meters. (b) 100 meters, (c) 300 meters, and (b) 500 meters.}\vspace{-4mm}
\label{fig:all_CDF}
\end{figure*}
\section{Numerical Results}
In this section, we present numerical results on the performance of the TDOA-based localization scheme for the four antenna patterns, using the CRLB expression derived in Section~\ref{sect:CRLB}. In future extension of this work, we intend to corroborate these analytical results with simulation results as well. We consider an area of size $1000{\times}1000\,{\text{m}^2}$ with the origin located at the origin $(0,0)\,{\text{m}}$, and $4$ RF sensors placed at $(250,250)\,{\text{m}}$, $({-}250,250)\,{\text{m}}$, $({-}250,{-}250)\,{\text{m}}$, and $(250,{-}250)\,{\text{m}}$. We also assume that the RF sensors are operating at the carrier frequency of $5.8$~GHz, and that the bandwidth and the transmit power of the drone are $10$~MHz and $20$~dBm, respectively~\cite{whitepaper1,whitepaper2}, and that the noise power spectral density (PSD) is ${-}174$~dBm/Hz~\cite{carlson2002communication}. 
\par In order to characterize the coverage performance provided by the different antenna patterns, we plot the CDFs in Fig.~\ref{fig:all_CDF} for corresponding CRLB-RMSE. 
In each subplot of Fig.~\ref{fig:all_CDF}, we provide two horizontal lines (i.e., in magenta and black), which represent the $80\%$ (i.e, $80$\%-coverage-error-bound) and $50\%$ (i.e., median) coverage based on RMSE with $F_{E}(e)\,{=}\,80\%$ and $F_{E}(e)\,{=}\,50\%$, respectively. In addition, the vertical line (i.e., in blue) represents the coverage based on $\delta\,{=}\,100$ (i.e., $100$-meters-error-bound-coverage).
From Fig.~\ref{fig:all_CDF}(a), we observe that at very low drone altitudes, such as $10$ meters, the VH antenna pattern yields the lowest median RMSE, the lowest $80$\%-coverage-error-bound, and the highest $100$-meters-error-bound-coverage\%. Thus the VH pattern provides the best localization performance in terms of both accuracy and area coverage, followed by the VV, and the HH pattern. We also observe that the VV, and the `Uniform' pattern have almost identical distributions of RMSE at very low altitudes. 
\par Next, from Fig.~\ref{fig:all_CDF}(b), we observe that for medium drone altitudes of about $100$ meters, the VV pattern yields the highest accuracy in terms of the median RMSE, and the highest quality of localization (QOL) in terms of the $80$\%-coverage-error-bound. However we also note that the VV and the HH CDFs intersect above the horizontal line, $F_{E}(e)=0.9$, indicating that if we measured the QOL in conjunction with a higher requirement for ACE, such as $Pr(E\le\delta)\ge 90$\%, then the HH pattern would provide a lower $90$\%-coverage-error-bound and thus higher QOL as compared to the VV, and the `Uniform' pattern. Fig.~\ref{fig:all_CDF}(b) also shows us that, the value of the $100$-meters-error-bound-coverage\% is the highest for the HH curve, i.e. the value of the ACE subject to an error tolerance of $100$ meters is maximized for the HH antenna pattern. 

On the other hand, if we considered the ACE metric to be subject to a lower error tolerance such as $\delta \le 50$ meters, then the maximum ACE will be obtained from the VV and the `Uniform' pattern, followed by the VH pattern, instead of the HH pattern. 
From Fig.~\ref{fig:all_CDF}(c), and (d), we observe that at very high drone altitudes, such as $300$ to $500$ meters,
the lowest median RMSE, the lowest $80$\%-coverage-error-bound, and the highest $100$-meters-error-bound-coverage\% are achieved by the `Uniform' pattern. While the performance of the VV pattern closely follows that of the `Uniform' pattern, we note that in the case of a very high ACE requirement ($Pr(E\leq\delta) \ge 90$\%), the the highest QOL is achieved by the VV pattern, and in the case of a higher error tolerance ($\delta \geq 150$ meters), the higher ACE is also provided by the VV pattern. Unlike, low and medium altitudes, at very high  altitudes, we see more differentiation between the RMSE distributions for the VV and the `Uniform' pattern, and the CDFs of the VH and the HH patterns move closer to that of the `Uniform' pattern. This can be explained by the fact that at lower altitudes the elevation angle is very small, thus the cosine of the elevation angle, and the antenna gain for the VV pattern is very close to the gain for the the `Uniform' pattern, $1$. 

As the altitude increases, the elevation angle becomes larger, resulting in lower antenna gain for the VV pattern, and higher antenna gain for the VH, and the HH pattern. Therefore the performance gap between the VV and the `Uniform' pattern becomes larger, while the performance gap between the HH/VH pattern and  the `Uniform' pattern becomes smaller.
\begin{figure}[t]\vspace{-1mm}
	\centering	
	\includegraphics[width=.8\columnwidth]{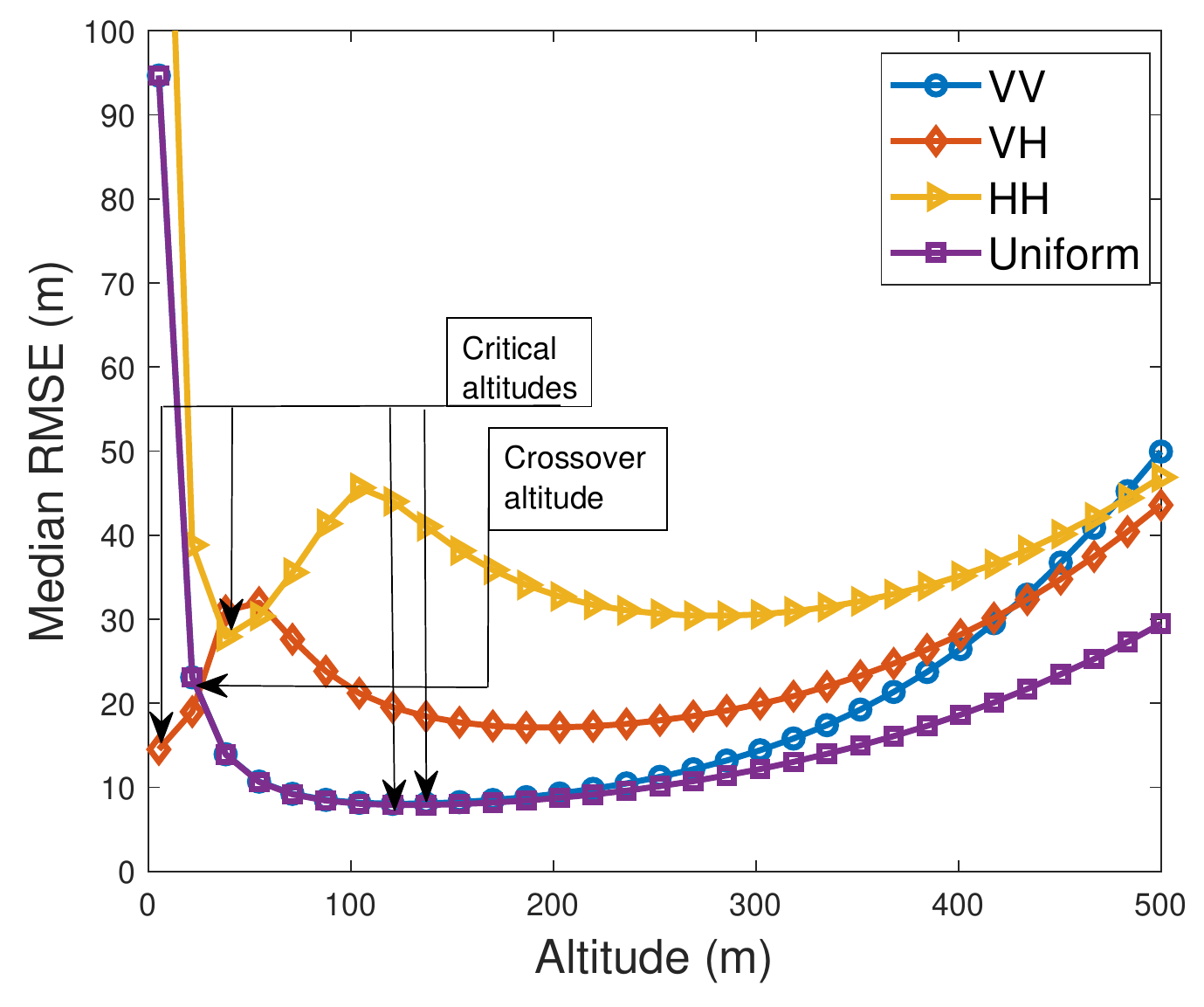}\vspace{-1mm}
	\caption{Median RMSE as a function of drone altitude for various antenna patterns using 4 sensors.}
	\label{fig:all_meadian_together}
    \vspace{-5mm}
\end{figure}
\begin{figure}[t]\vspace{-1mm}
	\centering	
	\includegraphics[width=.8\columnwidth]{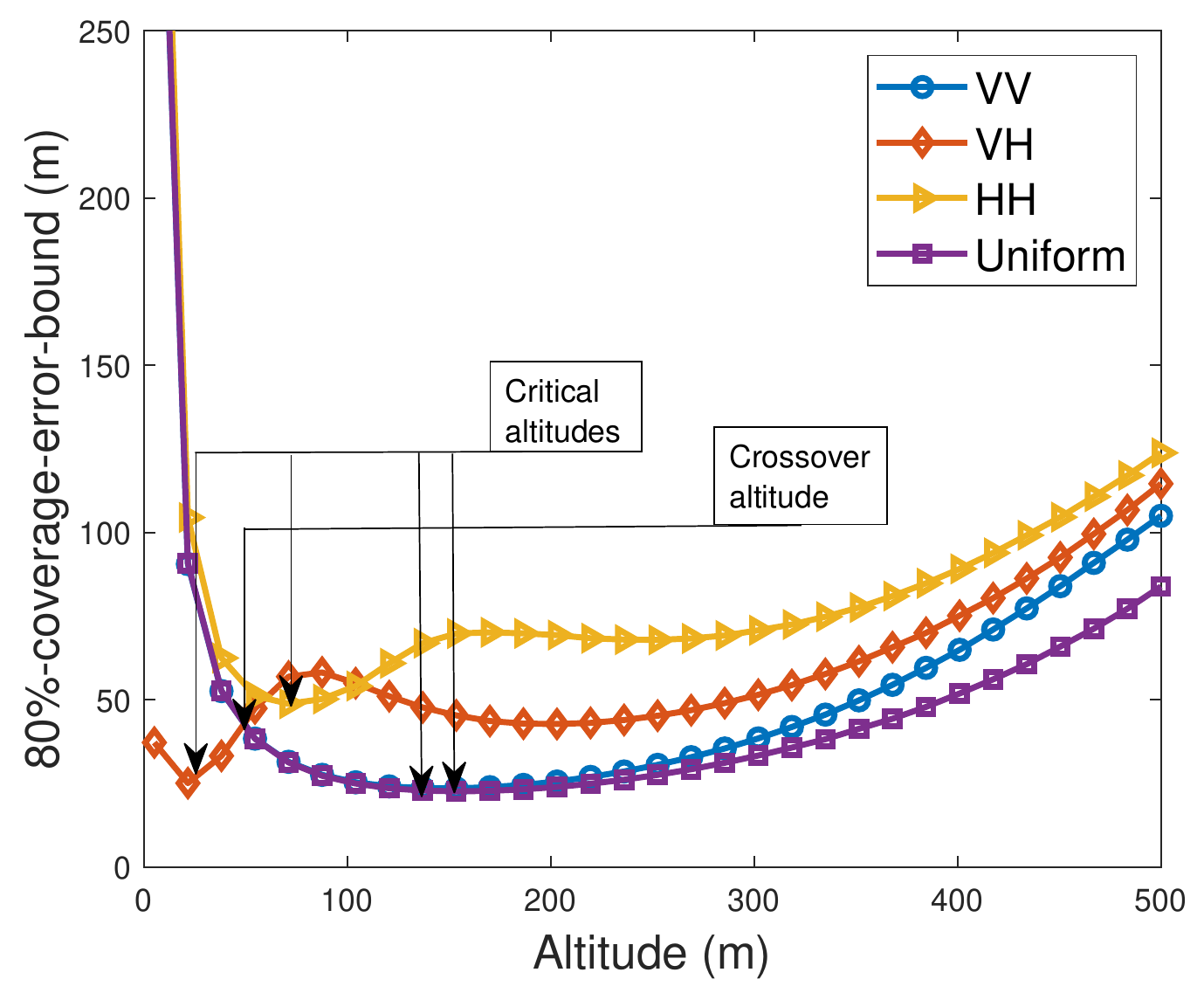}\vspace{-1mm}
	\caption{$80$\%-coverage-error-bound as a function of drone altitude for various antenna patterns using 4 sensors.}
	\label{fig:all_ept}
    \vspace{-4mm}
\end{figure}
\begin{figure}[t]\vspace{-1mm}
	\centering	
	\includegraphics[width=.8\columnwidth]{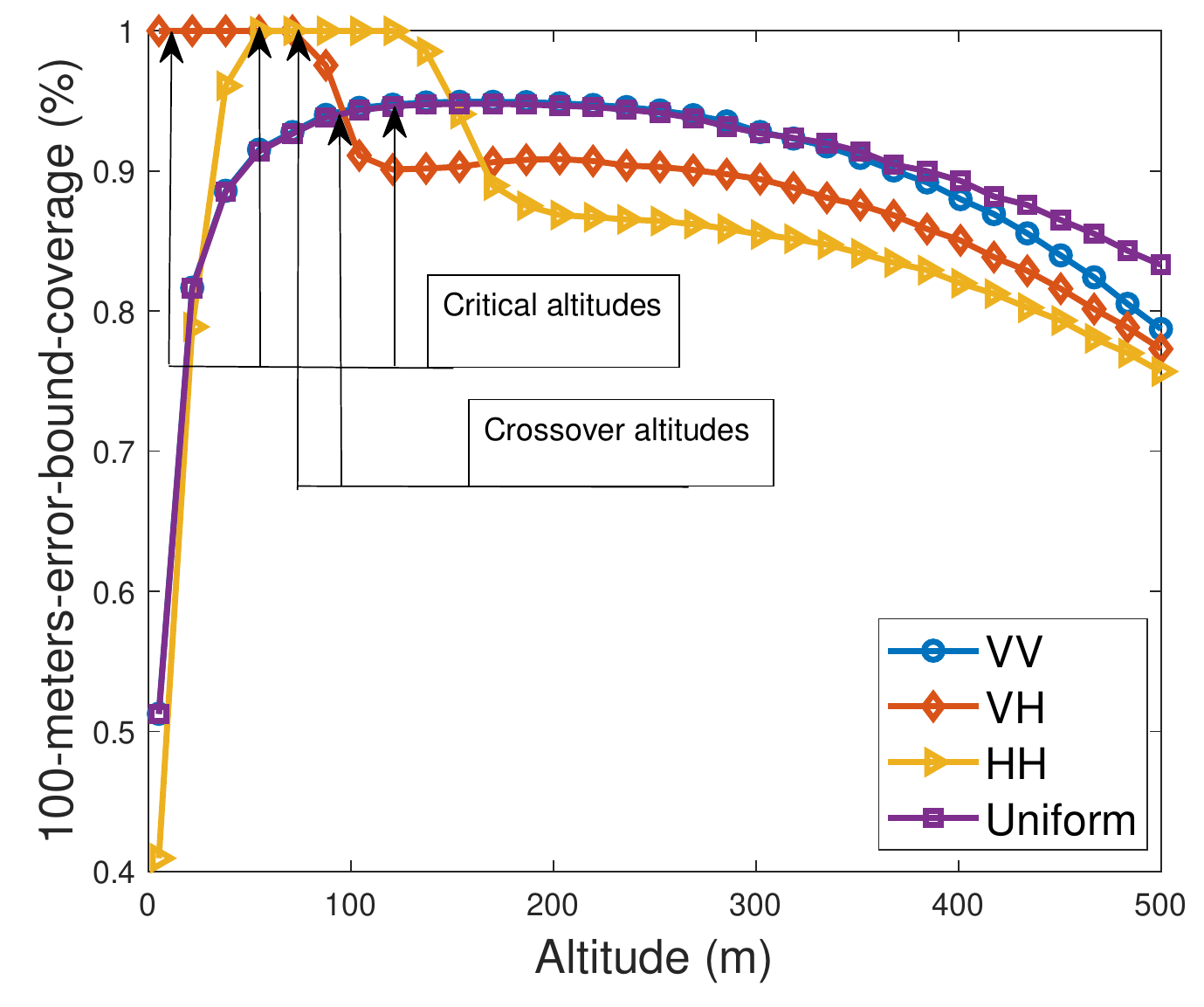}\vspace{-1mm}
	\caption{$100$-meters-error-bound-coverage\% as a function of drone altitude for various antenna patterns using 4 sensors.}
	\label{fig:all_hundc}
    \vspace{-5mm}
    \end{figure}

Having realized the importance of characterizing the localization performance with respect to the drone altitude, we plot these performance metrics as a function of the drone altitude, in Fig.~\ref{fig:all_meadian_together}, Fig.~\ref{fig:all_ept}, and Fig.~\ref{fig:all_hundc}. From these plots, we observe a general pattern, that is, at very low altitudes below $50$ meters, all antenna patterns other than the VH pattern, yield very high errors, therefore we observe very high median RMSE, and $80$\%-coverage-error-bound, and very low $100$-meters-error-bound-coverage\%. Beyond $50$ meters, the performance for these antenna patterns start to improve, and reaches the optimal point at a certain altitudes, but again starts to degrade at very high altitudes. Thus we realize, that there is a critical height where the performance metrics are optimized. For the VH pattern, the performance is optimized at very low altitudes such as below $50$ meters. The local maximas/minimas for the different metrics, have been annotated in the corresponding plots for all the antenna patterns. Table~\ref{Tab1} summarizes the results.


The points of intersection between the different antenna pattern curves, in Fig.~\ref{fig:all_meadian_together}, Fig.~\ref{fig:all_ept}, and Fig.~\ref{fig:all_hundc}, indicate that the choice of the best antenna pattern with respect to the given performance metric, changes before and after the corresponding point of intersection. The abscissa of the point of intersection gives us the crossover height, beyond which the choice of the best antenna pattern changes. These crossover altitudes have been annotated in the plots. We observe that, in terms of the median RMSE the crossover altitude is $21.5$ meters. Below $21.5$ meters the highest accuracy is achieved by the VH pattern, and beoynd $21.5$ meters the accuracy is maximized by the `Uniform' pattern. In terms of QOL, the lowest ept is provided by the VH pattern for drone altitudes below $55$ meters, beyond which the best QOL is achieved by the `Uniform' pattern. Lastly we note that, when ACE is measured in terms of the hundc, the best coverage performance is provided the VH pattern till the drone altitude of $71$ meters, whereas between $71$ meters and $121$ meters, the HH pattern, and beyond $121$ meters, the `Uniform' pattern, respectively, yield the highest ACE. 

\begin{table}[!t]
\vspace{1mm}
\centering
 \caption{Critical drone altitude with respect to various performance metrics for 4 sensors.}\label{Tab1}
 \begin{tabular}{|c||c|c|c|c|} 
 \hline  & {\bf VV} & {\bf VH} & {\bf HH}& {\bf Uniform} \\ 
 \hline  Median RMSE & $130$ & $5$ & $15$ & $45$\\ 
 \hline  $80$\%-coverage-error-bound & $145$ & $20$ & $75$ & $155$\\ 
 \hline  $100$-meters-error-bound-coverage\% & $160$ & $5$ & $55$ & $150$\\
 \hline
 \end{tabular}
\end{table}

\section{Conclusion}
\label{sect:ConclusionmmWave}
In this paper a TDOA-based RF positioning system for localization of drones is studied in conjunction with a simple air-to-ground 3D antenna radiation pattern.
Our results show that accounting for antenna effects makes a significant difference and reveals many important relationships between the localization accuracy and the altitude of the drone. It is seen that the localization performance varies in a non-monotonic pattern with respect to the drone altitude. We are also able to characterize, the critical heights, where the coverage and accuracy metrics are optimized, and the crossover UAV heights, before and after which, the choice of the best antenna pattern yielding best performance changes. In the future, we intend to extend this work, primarily by carrying out a similar analysis on the localization CRLB in conjunction with a more realistic antenna pattern, and then finding the optimal placement of the ground sensors, that minimizes the localization CRLB.
\bibliographystyle{IEEEtran}
\bibliography{refs}
\end{document}